\documentclass[fleqn,usenatbib]{mnras}
\usepackage{newtxtext,newtxmath}
\usepackage[utf8]{inputenc}
\usepackage[T1]{fontenc}

\usepackage{ae,aecompl}
\usepackage{graphicx}
\usepackage{amsmath}
\usepackage{amssymb}
\usepackage{xcolor}
\usepackage{soul}
\usepackage[normalem]{ulem}

\newcommand{\Emax}{E_\text{max}}
\newcommand{\Emaxl}{E_\text{max,l}}
\newcommand{\Emaxh}{E_\text{max,h}}
\newcommand{\alphal}{\alpha_\text{l}}
\newcommand{\alphah}{\alpha_\text{h}}
\newcommand{\Lc}{L_\text{c}}
\newcommand{\dtFlare}{\Delta t_\text{flare}}
\newcommand{\dtAgn}{\Delta t_\text{AGN}}
\newcommand{\dtIgmf}{\Delta t_\text{IGMF}}
\newcommand{\txs}{TXS~0506+056}

\definecolor{darkgreen}{rgb}{0.0, 0.5, 0.0}

\title[TXS 0506+056: Intrinsic $\gamma$-Ray Spectrum \& Propagation Effects]{
The Intrinsic Gamma-Ray Spectrum of TXS~0506+056: Intergalactic Propagation Effects}

\author[A. Saveliev \& R. Alves Batista]{Andrey Saveliev$^{1,2}$\thanks{andrey.saveliev@desy.de}, Rafael Alves Batista$^{3}$\thanks{r.batista@astro.ru.nl (corresponding author)}
\\
$^1$ Immanuel Kant Baltic Federal University, Institute of Physics, Mathematics and Information  Technology, 236041 Kaliningrad, Russia \\
$^2$ Lomonosov Moscow State University, Faculty of Computational Mathematics and Cybernetics, 119991 Moscow, Russia \\
$^3$ Radboud University Nijmegen, Department of Astrophysics/IMAPP, 6500 GL Nijmegen, The Netherlands
}

\date{Accepted 2020 October 28. Received 2020 October 14; in original form 2020 September 3}

\pubyear{2021}

\begin{document}
\label{firstpage}
\pagerange{\pageref{firstpage}--\pageref{lastpage}}
\maketitle

\begin{abstract}
The recent observation of high-energy neutrinos from the 2017 flare of the blazar TXS~0506+056, together with counterparts across the whole electromagnetic spectrum, opens up new possibilities for investigating the properties of this class of objects as well as the traversed medium. Propagation effects such as the attenuation of the very-high-energy gamma-ray component by the extragalactic background light are well known, and usually taken into account when fitting spectral energy distributions of objects. Other effects such as those of intergalactic magnetic fields are, however, often neglected. In this work, we present a comprehensive study of the influence of these fields and the extragalactic background light on the determination of the intrinsic gamma-ray spectrum of this blazar.
\end{abstract}


\section{Introduction}\label{sec:intro}

Active galactic nuclei (AGNs) are sources of electromagnetic radiation across the whole spectrum, from radio to gamma rays~(for a review, see, e.g., \citealt{padovani2017a}). Blazars are a type of AGN with relativistic jets pointing approximately towards Earth, making them interesting objects to study the extreme processes taking place near accreting supermassive black holes at their centres. They have been known to be sources of energetic gamma rays for decades~\citep{punch1992a}. 

The high-energy (HE; $\gtrsim 1 \; \text{GeV}$) and very-high-energy (VHE; $\gtrsim 100 \; \text{GeV}$) electromagnetic component of the spectral energy distribution (SED) of AGNs is sometimes attributed to leptonic processes~\citep{dermer1992a,schlickeiser1996a,mastichiadis1997a,ghisellini1998a,boettcher2013a,diltz2014a}, involving interactions of electrons with photon and magnetic fields pervading the environment. The emission by these objects can also be due to hadronic processes (see e.g.~\citealt{mannheim1992a,mannheim1995a,mastichiadis1996a,halzen1997a,cerruti2015a,tavecchio2015a,khiali2016a,murase2016a,zech2017a}), where HE cosmic rays interact with the environment producing pions, whose decays create very energetic gamma rays and neutrinos. Lepto-hadronic models also exist (see, e.g., \citealt{muecke2001a,cerruti2019a,rodrigues2019a}). They are commonly fits to the observed multiwavelength SED based on single or multiple zones. Blazars or, more specifically, BL Lac objects, have no strong and broad lines in their optical spectrum. A typical SED for this class of objects features two humps, and they are usually classified by the energy of the first of these peaks \citep{Padovani:1994sh}, which is mostly due to synchrotron emission by relativistic electrons and positrons within the blazar's environment. A possible subdivision is given in \citet{Abdo:2009iq}, which distinguishes between (a) low synchrotron peaked (LSP) blazars, which peak in the (far) infrared band, i.e.~at peak frequencies $f_{\rm peak} \lesssim 10^{14}\,{\rm Hz}$;  (b) intermediate synchrotron peaked (ISP) blazars with $10^{14}\,{\rm Hz} \lesssim f_{\rm peak} \lesssim 10^{15}\,{\rm Hz}$; and (c) high synchrotron peaked (HSP) blazars with $f_{\rm peak} \gtrsim 10^{15}\,{\rm Hz}$, such that the peak energy lies in the UV or X-ray region. The HE peak can be explained by inverse Compton scattering of low-energy background photons by electrons, by synchrotron emission due to highly relativistic protons, or by photopion production of HE photons; see \citet{boettcher2013a} for more details. Naturally, combinations of these processes are also plausible explanations. 

It is difficult to ascribe a purely hadronic/leptonic interpretation to the VHE emission by blazars based solely on gamma rays. For this reason, in addition to multiwavelength observations, neutrino measurements are  important, since they allow us to distinguish among the aforementioned scenarios. 

The observation of the high-energy neutrino event IC~170922A ($E_\nu \simeq 290 \; \text{TeV}$) correlating with the position of the blazar \txs~\citep{icecube2018b}, in combination with an electromagnetic counterpart~\citep{icecube2018b}, was the cornerstone of multimessenger astronomy. \txs{} is a laboratory for testing models of non-thermal emission by blazars. HE emission from this object had already been observed by EGRET~\citep{dingus2001a}. Located at $z \simeq0.3365$~\citep{paiano2018a}, \txs{} also had another (even stronger) episode of neutrino emission during 2014-2015.  A total of $13 \pm 5$ neutrinos were detected within a time window of 110~d~\citep{icecube2018a}. Archival analyses of {\it Fermi}-LAT data for this period reveal that \txs{} was at a low state during this period~\citep{garrappa2019a}. The scarcity of multiwavelength observations make it difficult to draw a picture of this object capable of accommodating both the 2014-2015 and the 2017 flares.

Hadronic and lepto-hadronic models were put forth to interpret the observations of \txs~\citep{magic2018a,keivani2018a,sahakyan2018a,Samui:2017dsz,cerruti2019a,gao2019a,liu2019a,rodrigues2019a}. They infer a maximum cosmic-ray energy of $E_\text{CR} \sim 10^{15} - 10^{18} \; \text{eV}$. Following the common procedure of SED-fitting, the intrinsic parameters of the object were constrained. One step in this procedure is to include the opacity of the Universe to pair production ($\tau_{\gamma\gamma}$), whose dominant contribution corresponds to infrared photons from the extragalactic background light (EBL). This photon field effectively leads to an exponential attenuation of the flux, of the form $\exp(-\tau_{\gamma\gamma})$. An issue that arises naturally is: \emph{If VHE gamma rays create electron-positron pairs in the intergalactic medium, what happens to these charged particles in the presence of intergalactic magnetic fields?}

The question posited before relates to a yet unanswered problem in cosmology: the origin of the magnetic fields in the Universe. Currently, two general classes of magnetogenesis scenarios are considered in the literature: cosmological and astrophysical. For the former, one assumes that strong seed magnetic fields have been created in the very early Universe (e.g.~during the electroweak or QCD phase transition, or during inflation -- see e.g.~\citealt{durrer2013a} and \citealt{Subramanian:2015lua}) and then evolve up to the present day, which may be simulated in full or (semi-)analytical magnetohydrodynamic simulations \citep{Saveliev:2012ea,Saveliev:2013uva,Kahniashvili:2012uj,Campanelli:2013iaa,Kahniashvili:2015msa,Kahniashvili:2016bkp,Brandenburg:2020vwp}. On the other hand, the basic idea of the astrophysical scenario is that weak magnetic field seeds were created during later stages of the evolution of the Universe (and then amplified by a battery mechanism), e.g.~during Reionization \citep{Langer:2018bbk}, from cosmic-ray currents \citep{Miniati:2010ne,Ohira:2020yvb}, or galactic (nuclei) outflows \citep{Furlanetto:2001gx,Beck:2012cs,Samui:2017dsz}.

Lower limits on the strength of IGMFs were obtained by several authors \citep{neronov2010a,Tavecchio:2010mk,Tavecchio:2010ja,Dolag:2010ni,Essey:2010nd,Finke:2015ona,Biteau:2018tmv,saveliev2020b}
using the deflection of electron-positron pairs produced in electromagnetic cascades in IGMFs. An immediate consequence of the existence of such fields is an angular spread of the arrival directions of secondary gamma rays produced via inverse Compton scattering (see e.g~\citealt{Dolag:2009iv,neronov2010b,Chen:2014rsa,Yang:2015lqy}). Evidently, temporal profiles are also expected to depend on properties of the fields \citep{plaga1995a,Murase:2008pe,Oikonomou:2014xea,Yang:2015lqy}. For this effect, however, the duration of the emission has to be comparable to the other time-scales involved (observation time). Another observable that is directly influenced by IGMFs is the flux of gamma rays observed at Earth \citep{dAvezac:2007xri,neronov2010a,vovk2012a}. For longer time delays, part of the HE band of the spectrum cannot be detected, as not all photons can arrive within a fixed time window. Moreover, the more the electrons (and positrons) in the cascade are deflected, the more diluted the arrival distribution of photons from a single source is, and the less likely it is that they will be contained within a fixed angular window corresponding to the object. Consequently, IGMFs induce a flux suppression of secondary photons; but they are \emph{not} accounted for in SED fits. It should be noted here that there is another explanation for this suppression: Electrons/positrons present in the cascade might interact with the intergalactic medium and generate plasma instabilities, leading to energy losses and, consequently, to a lower flux of HE photons arriving at the observer \citep{broderick2012a,2012ApJ...758..102S,Miniati:2012ge,Schlickeiser:2013eca,Sironi:2013qfa,broderick2018a,Vafin:2018kox,Yan:2018pca,alvesbatista2019a} . 

In light of the preceding discussion, in this work, we address how the presence of intervening IGMFs may interfere with the modelling of the intrinsic spectral parameters of blazars. We focus on the particular case of the blazar \txs. We start off by describing the three-dimensional simulation set-up adopted, in Section~\ref{sec:simulation}, and then we explain the procedure to post-process these simulations in Section~\ref{sec:analysis}. In~\ref{sec:first} we present the results of a first intuition-builder study required for the interpretation of the results. The procedural details concerning the fitting of the observations is extensively addressed in Section~\ref{sec:fit}, followed by the results in~\ref{sec:results}. Finally, in Section~\ref{sec:discussion}, we discuss our results, and draw our conclusions in~\ref{sec:conclusions}.

\section{Simulation setup}\label{sec:simulation}

We simulate the propagation of gamma rays in the intergalactic space using the {\sc crpropa} code~\citep{alvesbatista2016b}. It is a modular code designed for the propagation of HE particles in the Universe. We consider pair production ($\gamma_\text{HE} + \gamma_\text{bg} \rightarrow e^{+} + e^{-}$) and inverse Compton scattering ($e^\pm + \gamma_\text{bg} \rightarrow e^{\pm} + \gamma_\text{HE}$) both in the EBL and CMB; here the subscript `HE' denotes a high-energy particle, and `bg' a background photon. Adiabatic energy losses due to the expansion of the Universe are considered as well. In this case, a particle with initial energy $E_{0}$ would be observed with an energy $E(z) = E_0 / (1 + z)$, where $z$ is the redshift of emission. We also consider synchrotron radiation by charged particles; this is, however, small, and produce photons below the energy range of interest ($\ll 1 \; \text{GeV}$). 

The source is assumed to be located at the centre of a sphere of radius $D$, which corresponds to the (co-moving) distance from the Earth to \txs{}, following the large sphere observer method described in detail in \citet{alvesbatista2016a}. Events are emitted isotropically until one of the following conditions are met: (i) They hit the sphere; (ii) their energy drops below a minimum energy threshold, assumed here to be $1 \; \text{GeV}$; (iii) and the total trajectory length described by the particle exceeds a maximum threshold of 4000~Mpc. Each event that hits the sphere represents a particle arriving at Earth. The line of sight is represented by the original direction of emission of the particle, and the direction where the arriving particle crosses the sphere is its arrival direction, assuming a coordinate system centred at the source (centre of the sphere). The total flux is obtained by integrating over all contributions of individual particles, with the cuts described in section \ref{sec:analysis}.

We use the built-in {\sc crpropa} integrator, which solves the equations of motion with a fifth-order Runge-Kutta method with adaptive steps. The minimum and maximum step sizes are, respectively, $10^{13} \; \text{m}$ and $10 \; \text{Mpc}$. This choice yields a time resolution of $\sim 1 \; \text{d}$.

The magnetic field is assumed to be a turbulent  zero-mean Gaussian random field with a Kolmogorov spectrum, and redshift evolution given by $B(z) = B_0  (1 + z)^2$. It is sampled in the Fourier space, transformed into real space, and projected onto a uniformly spaced cubic grid with $N = 500^3$ cells. The minimum scale that can be resolved is $\ell_\text{min} \equiv 2 L_{\rm g} / N^{1/3}$, following the Nyquist criterion, where the grid size ($L_\text{g}$) is $L_\text{g} = 20 \; \text{Mpc}$ for $\Lc \leq 1 \; \text{Mpc}$, and $L_\text{g} = 10\Lc$ otherwise. The maximum scale is chosen to achieve the desired coherence length. The grid is periodically repeated to cover the whole volume between \txs{} and Earth. At each run, we change the seed used to generate the magnetic field grids to prevent spurious features inherent to one realization of the field from affecting the results. 

The magnetic fields considered range from $10^{-19}$ up to $10^{-14} \; \text{G}$, in logarithmic steps of 1. The coherence length ($\Lc$) lies in the range $10 \; \text{kpc}$--$1 \; \text{Gpc}$, also in logarithmic steps of 1. In addition, we consider the case $B=0$. 
The range of $B$ we adopt covers the typical lower bounds derived using electromagnetic cascades. We could not simulate stronger magnetic fields because the Larmor radii of electrons start to become excessively small, rendering it impractical to track single particles due to the high computational load -- especially if they can get trapped within a small region of space. The values of coherence length were chosen to encompass the most common values of $\Lc$ according to various constraints (see ~\citealt{durrer2013a} for a review). 

\begin{figure}
  \includegraphics[width=.98\columnwidth]{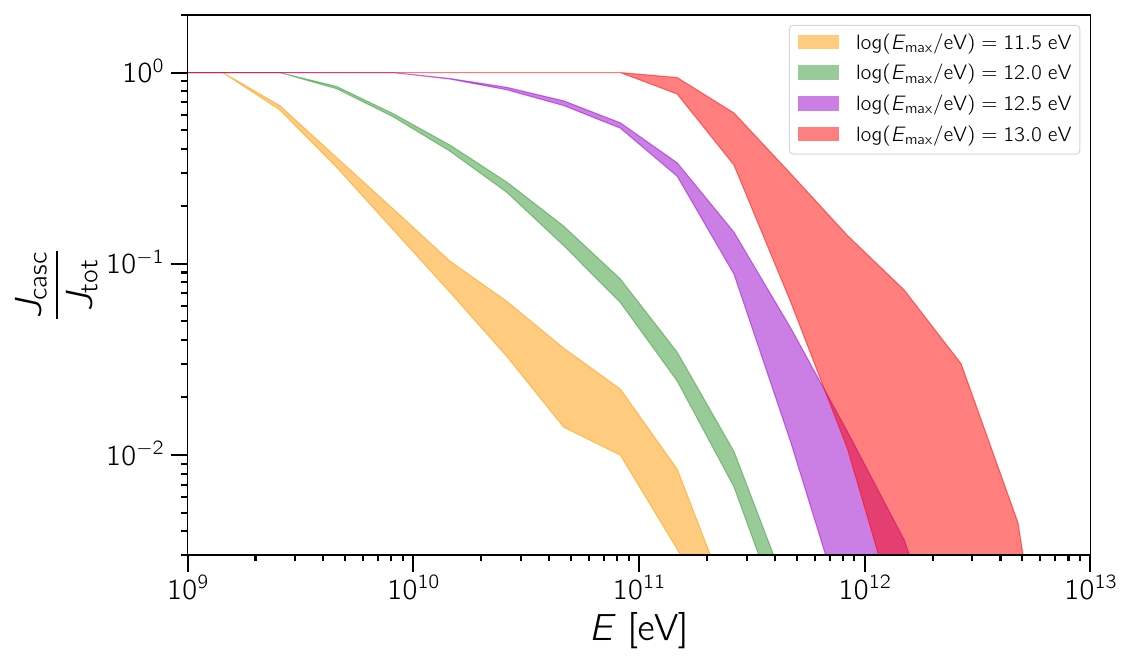}
  \caption{Contribution of cascade photons ($J_\text{casc}$) to the total flux ($J_\text{tot}$) at different energies, for different values of $\Emax$. The shaded regions represent the interval spanned by the four EBL models considered: \citet{dominguez2011a}, \citet{gilmore2012a}, and the upper and lower limits by \citet{stecker2016a}.}
  \label{fig:cascade}
\end{figure}

\begin{figure*}
	\includegraphics[width=\columnwidth]{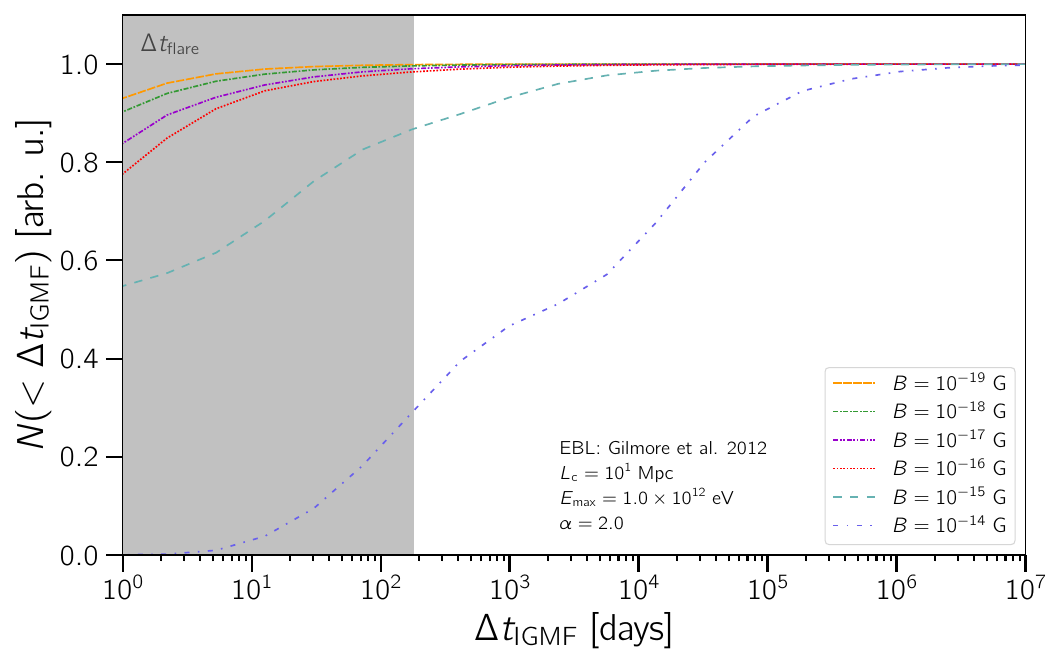}
	\includegraphics[width=\columnwidth]{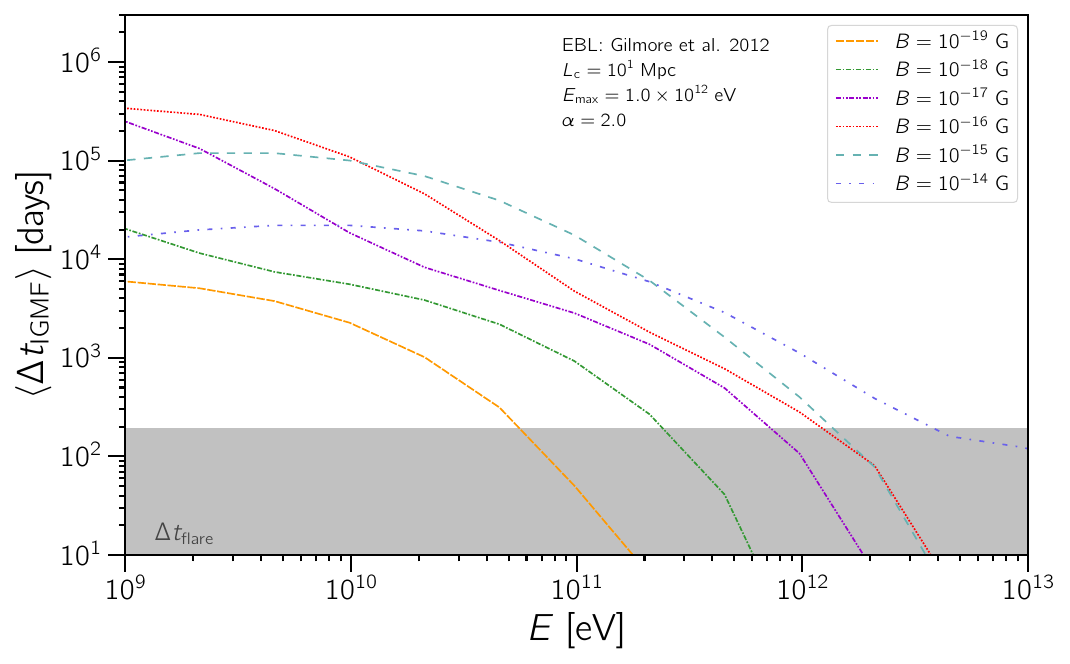}
	\caption{Cumulative distribution of time delays of arrival times (left-hand panel) and average time delays as a function of the observed energy (right-hand panel). The grey shaded region indicate the period of enhanced activity of the object ($\dtFlare \approx 180 \; \text{d}$). This particular figure assumes a spectral index $\alpha=2.0$ and $\Emax = 1 \; \text{TeV}$. These plots are for the whole energy range considered in the simulation ($E \ge 1 \; \text{GeV}$).}
	\label{fig:dtdist}
\end{figure*}

\section{Post-processing of the simulations}\label{sec:analysis}

We reject all events whose time delays exceed $\Delta t$ to compute the gamma-ray fluxes at Earth. For the enhanced emission state, we set $\Delta t = \dtFlare = 6 \; \text{months}$, which corresponds to the duration of the neutrino emission \citep{icecube2018b}. The duration of gamma-ray activity of \txs{} during its quiescent state is unknown, so we use $\Delta t = \dtAgn = 10, 10^{4}, \, 10^{7}$~yr.

We apply a posteriori cuts to the simulations in order to mimic the desired injection spectrum. This is done by applying weights to each simulated event according to the desired injection spectrum. Note that in our analysis, we are not concerned with the gamma-ray spectrum within the source environment. Instead, we consider only the gamma rays that escape the source; these two quantities are not necessarily the same. For a spectrum ${\rm d}N/{\rm d}E \propto E^{-\alpha} \exp(- E / \Emax)$, we consider the following parameter range: $ 0 \leq \alpha \leq 4$ and cut-off energy $10^{10.6} \leq \Emax / \text{eV} \leq 10^{14}$, in steps of $0.2$ for $\alpha$ and $\log(\Emax / \text{eV})$. 

The observed flux depends on the spectrum \emph{at the source}, which has two components: one for the low state (subscript `l'), and another for the enhanced emission (`h') periods. So, the emitted spectrum of gamma rays can be written as
\begin{equation}
	\frac{dN}{dE} \propto \left[ E^{-\alphal} \exp\left( -\frac{E}{\Emaxl} \right) + \eta E^{-\alphah} \exp\left( -\frac{E}{\Emaxh} \right) \right],
	\label{eq:spec}
\end{equation}
wherein $\eta$ denotes the flux enhancement in the high state with respect to the low state. Note that this phenomenological model refers to the spectrum emitted by the source, which comprises two components that, together, lead to the flux measured at Earth within an $\sim$six-month time window.

We consider \txs{} to have a perfectly collimated jet pointing directly to Earth. While the viewing and jet opening angles could impact our predictions quantitatively, our arguments still hold.

TXS 0506+056 is a point-like source. Therefore, before confronting our simulations with the data, we have to fold in the point spread function of the detectors. These are introduced as energy-dependent windows around the position of the blazar. For energies below $300 \; \text{GeV}$, in the range of {\it Fermi}-LAT observations, we consider a 68\% containment radius and use the values from the instrument response function \texttt{P8R2\_V6}. For $E > 300 \; \text{GeV}$, we assume an angular resolution of $0.1^{\circ}$ up to $E = 1 \; \text{TeV}$ and $0.06^{\circ}$ above this energy; these are similar to MAGIC's angular resolution~\citep{magic2016a}. 

In the simulations there are no background events contributing to the flux within the containment radius. However, simulated events whose angular distances to the source position exceed the size of the window associated with the containment radius have to be rejected to ensure consistency with the observations~\citep{icecube2018a,icecube2018b,magic2018a}. This is important to guarantee that the significance of the correlation between the gamma-ray signals and the electromagnetic/neutrino counterparts is preserved.

\section{A first study}\label{sec:first}

Before we proceed and fit the observations, it is in order to provide further grounds for our forthcoming results. 

The kinematic energy threshold for the production of pairs by HE gamma rays is $E_\text{thr} \sim m_e^2 / \varepsilon$, where $\varepsilon$ denotes the energy of the target photon. For the EBL, $\varepsilon \sim 0.001 - 10 \; \text{eV}$. Thus, from purely kinematic arguments, we expect that at least a fraction of the gamma rays from \txs{} be comprised of secondaries produced in the electromagnetic cascade process.

The EBL attenuates the HE component of the VHE emission, leading to a spectral suppression. This effect could, at first sight, be easily mistaken by a cut-off due to the maximum energy ($\Emax$) of gamma rays leaving the object. A natural question that then arises is: \emph{How can we distinguish between these two mechanisms?} If $\Emax \lesssim 100 \; \text{GeV}$, most of the observed flux is due to prompt emission. However, if $\Emax \gtrsim 400 \; \text{GeV}$, cascade photons could significantly contribute to the gamma-ray flux above 1~GeV.
The cut-off energy is usually determined through a two-step process. First, the intrinsic spectrum is inferred based solely on the lower energy component of the observed flux. Then a scan is performed to search for the value of $\Emax$ (the maximum energy emitted by the source) that best describes the data, including also the HE component and EBL absorption (see, e.g., \citealt{fermi2012a,dwek2013a,biteau2015a}). In this paper, we entertain the possibility  that some additional effect (magnetic fields) may influence the estimates of $\Emax$.

In this section, to better understand the problem and build our intuition for subsequent discussions, we adopt a simplified phenomenological model with only one of the components of equation~\ref{eq:spec}. Essentially, in this elementary model, we have a source at the same location as \txs{} whose activity leads to the emission of gamma rays during a time window of $\dtAgn$.

First, we prove that a cascade component from \txs{} is expected. We use the simulations described previously and neglect magnetic field effects. We estimate the ratio between the flux of secondary photons ($J_\text{casc}$), produced in the electromagnetic cascade, with the total flux ($J_\text{tot}$), which do not undergo the cascading process. 
These results are shown in fig.~\ref{fig:cascade}.

The highest energy data point for the gamma-ray flux, as observed by MAGIC~\citep{magic2018a}, is around $E \simeq 400 \; \text{GeV}$. From figure~\ref{fig:cascade}, we are led to conclude that, for this simplified model, a significant contribution of secondary photons produced in the electromagnetic cascade is expected for $\Emax \gtrsim 300 \; \text{GeV}$. For the two-component phenomenological model we employ in the subsequent fit, the reasoning is similar albeit not straightforward, as the fraction of cascade photons will be determined by an interplay of all spectral parameters ($\Emaxh$, $\Emaxl$, $\alphah$, $\alphal$, $\eta$).   

We now proceed with the inclusion of IGMFs in the simulations. One of its most evident effects on HE gamma rays is the time delay incurred by the field on the charged component of the cascade. For nearly steady sources, whose HE emission takes place over time-scales of tens of thousands of years or more, this effect is small. For shorter transient events, the flux of gamma rays at $E \gtrsim 1 \; \text{GeV}$ depends on the duration of the emission and properties of the intervening magnetic fields. In the case of \txs, we considered that the blazar was active over a time $\dtAgn$, in addition to the much shorter period of enhanced activity coinciding with the neutrino flare ($\dtFlare$). In fact, as mentioned above, \txs{} may be used to constrain the magnetic field strength $B$ and even its coherence length $\Lc$ \citep{saveliev2020b}.

The impact of IGMFs on the gamma-ray fluxes is more pronounced if the time delays due to the magnetic fields ($\dtIgmf$) are much larger than the duration of the flare, i.e., if $\dtIgmf \gg \dtFlare$. In figure~\ref{fig:dtdist} we illustrate this effect by showing the cumulative time-delay distribution for one specific scenario (left-hand panel). An energy-dependent version of the average time delays for different (observed) energies is shown in the right-hand panel.

Figure~\ref{fig:dtdist} suggests that for strong enough magnetic fields, the total flux is diluted over a large period of time (cf.~\citet{plaga1995a,Murase:2008pe,takahashi2008a,neronov2009a}). In other words, if the cumulative distributions reach~$\sim 1$ within the shaded region corresponding to the neutrino flare, then the effects of IGMFs are small and so is the flux suppression due to the field. This general behaviour holds qualitatively for all EBL models studied. Moreover, the dependence of this effect on $\dtAgn$ is small. 

\section{Fitting the data using the simulations}\label{sec:fit}

In order to perform the analysis described in this work, a stringent multistage data analysis is necessary. To this end, we use the simulated spectra, which comprises both prompt emission and cascade photons, according to the phenomenological model from eq.~\ref{eq:spec}. First, we determined the best fit for the low state. The only parameter in this fit is the overall normalisation factor ($\nu$), which scales the simulated spectrum ($J_{0}(E) = E {\rm d}N_{\rm sim}/{\rm d}E$) to fit the observations ($J_\text{obs} = \nu J_0$) for a spectrum $J_{0}(E) = E {\rm d}N_{\rm sim}/{\rm d}E$ resulting from a simulation, the fitted spectrum is given by $J(E) = \nu J_{0}(E)$. To calculate the optimal $\nu$, we calculate the normalized log-likelihood $\ln\mathcal{L}$ of the fit using the data points $(E_{i}, J_{i})$ with the formula
\begin{equation} \label{eq:lnL}
    \ln \mathcal{L} = - \sum_{i=1}^{n} \frac{(J_\text{sim}(E_{i}) - J_\text{obs}(E_i))^{2}}{2\sigma(E_{i})}\,,
\end{equation}
where $\sigma(E_{i})$ is given by
\begin{equation}
\sigma(E_{i}) =
\begin{cases}
\sigma_{i}^{+} \;\;\;\; {\rm if } \;\; J_0(E_{i}) \ge J_{i}\,, \\
\sigma_{i}^{-} \;\;\;\; {\rm if } \;\; J_0(E_{i}) < J_{i}
\end{cases}
\end{equation}
for which $\sigma_{i}^{-}$ and $\sigma_{i}^{+}$ are, respectively, the lower and upper standard deviations for the $i$th point. In other words, we consider the split normal distribution as the asymmetric generalization of the normal distribution. Now, in order to find the value of $\nu$ that fits the data, $\ln\mathcal{L}$ has to be maximized, which is done numerically for each simulation.

Next, we reduce the size of the parameter space to be scanned by reducing the number of scenarios to be analysed. To this end, we exclude scenarios whose best fits describe the data poorly, i.e., scenarios whose low-state fits result in $P$-values smaller than $10^{-3}$. Consequently, we retain some plausible scenarios whilst significantly reducing the size of the parameter space. Owing to the complex form of the likelihood function, the $P$-value for a given log-likelihood cannot be calculated analytically, and hence has been derived using Monte Carlo methods.

\begin{figure}
	\centering
	\includegraphics[width=.48\textwidth]{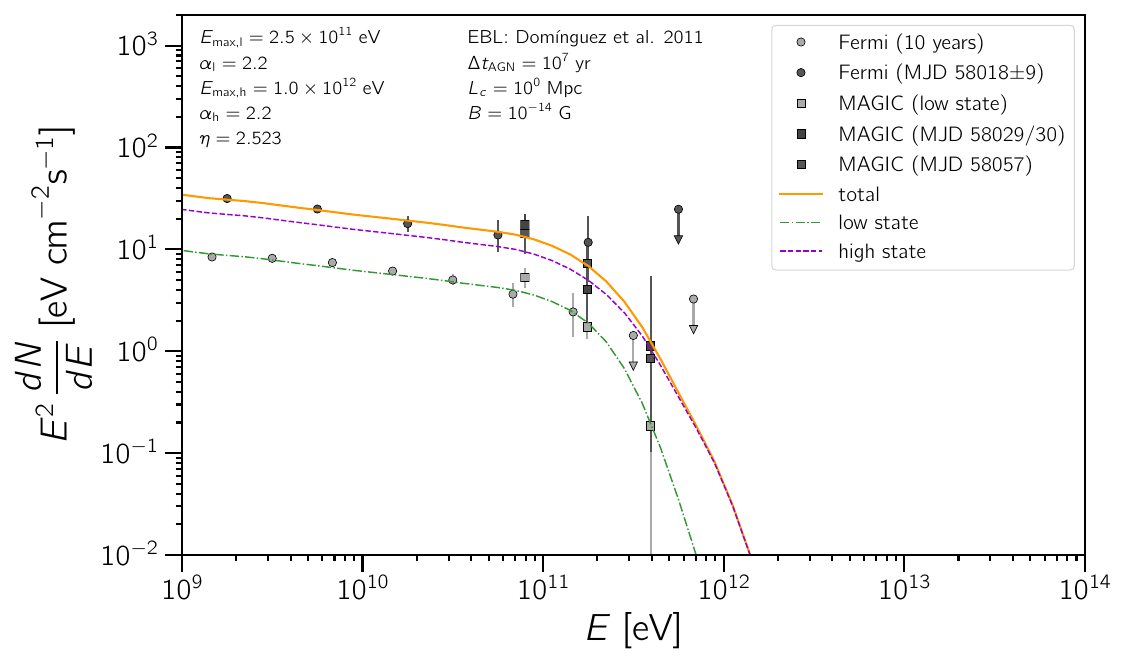}
	\caption{An example of the fit of the data with the results of the Monte Carlo simulations. The dot--dashed green line represents the fit to the low state. The dashed purple line corresponds to the high state ignoring the contribution of the low state. The combined total (low$+$high state) is shown as a thick orange solid line. Here, the EBL model from \citet{dominguez2011a} as well as the magnetic field parameters $B=10^{-14} \; \text{G}$ and $\Lc=1 \; \text{Mpc}$ were used.}
	\label{fig:spec}
\end{figure}

Now, for each low-state scenario selected in the previous step, we consider all possible scenarios with enhanced emission (i.e.~the simulations with the same $\Lc$ and $B$ as for the low state). For each of these combinations, we calculate the log-likelihoods, again using eq.~(\ref{eq:lnL}); this time, however, the $E_{i}$ denotes the energy values of the low state and the spectrum $J$ given by (\ref{eq:spec}). Here, $\eta$ is the parameter for which the best-fit value may be found by numerically maximizing the log-likelihood with respect to it. In fig.~\ref{fig:spec} we present a sample plot of a fit.

\section{Results}\label{sec:results}

In section~\ref{sec:first} we discussed the effects of IGMFs on the time distribution of gamma rays arriving at Earth. Building upon the discussion initiated there, we now present the results of the detailed fit of the observations for the two-state phenomenological model whose injection spectrum is given by equation~\ref{eq:spec}. It is worth stressing, once again, that our model concerns only gamma rays that escaped the source environment; they will propagate exclusively in the intergalactic medium.

\begin{figure*}
  \includegraphics[width=\columnwidth]{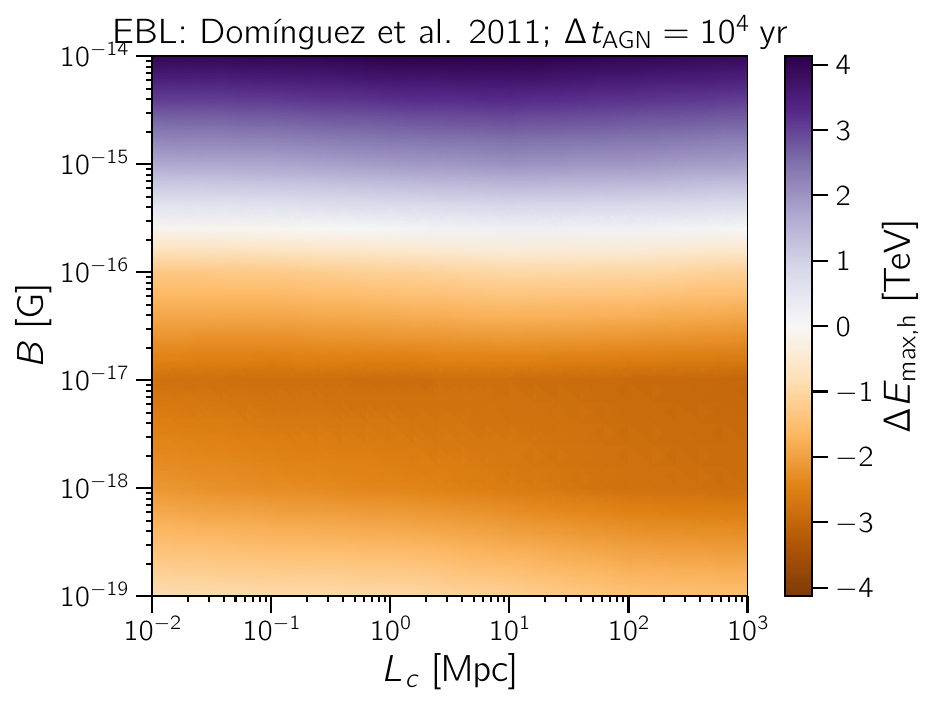}
  \includegraphics[width=\columnwidth]{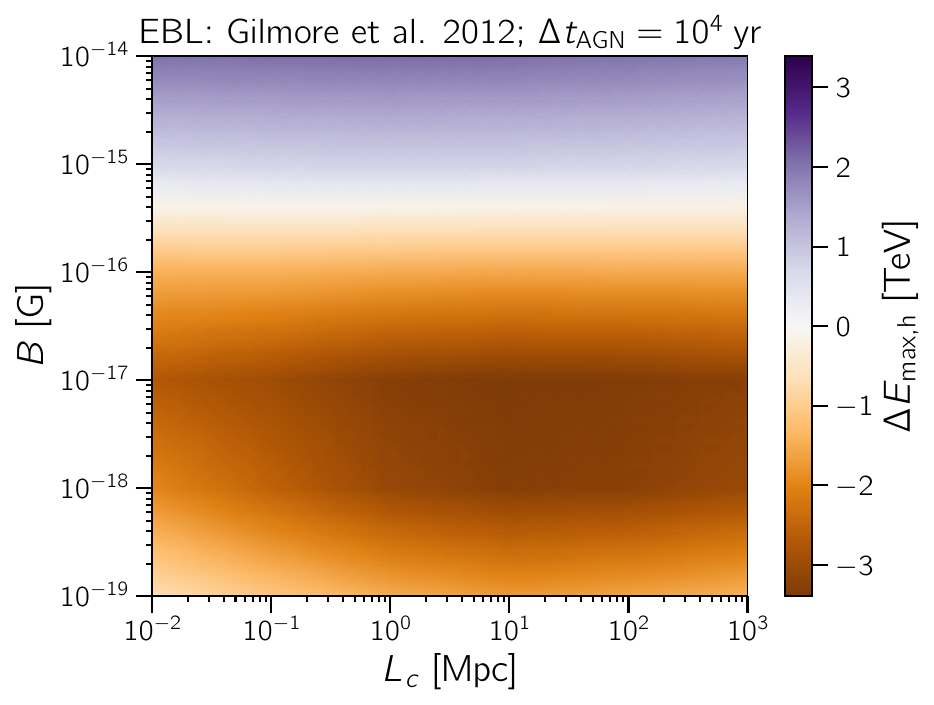}
  \includegraphics[width=\columnwidth]{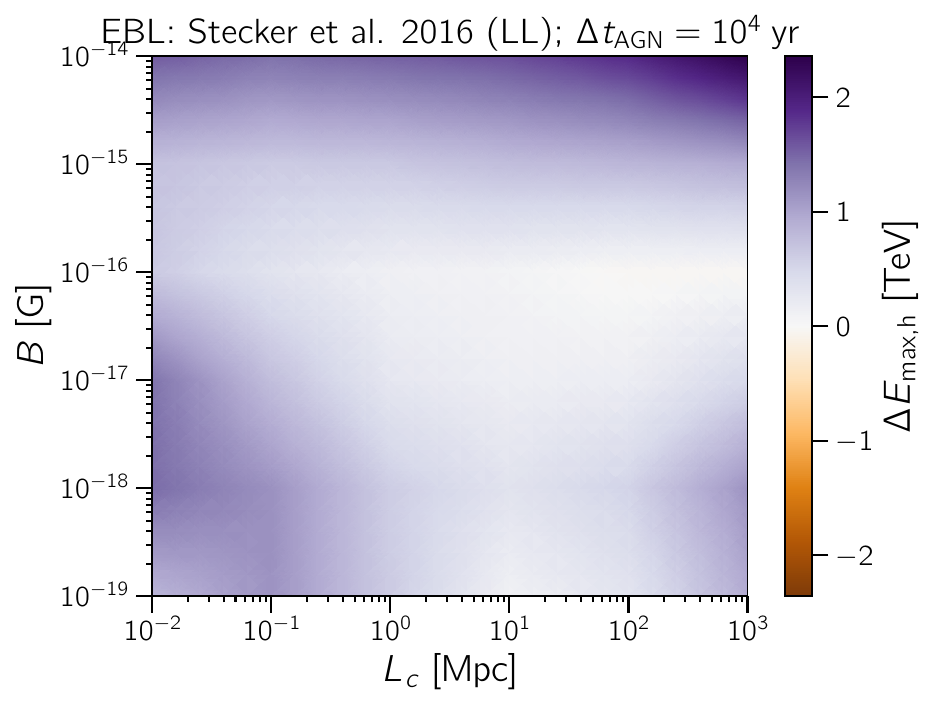}
  \includegraphics[width=\columnwidth]{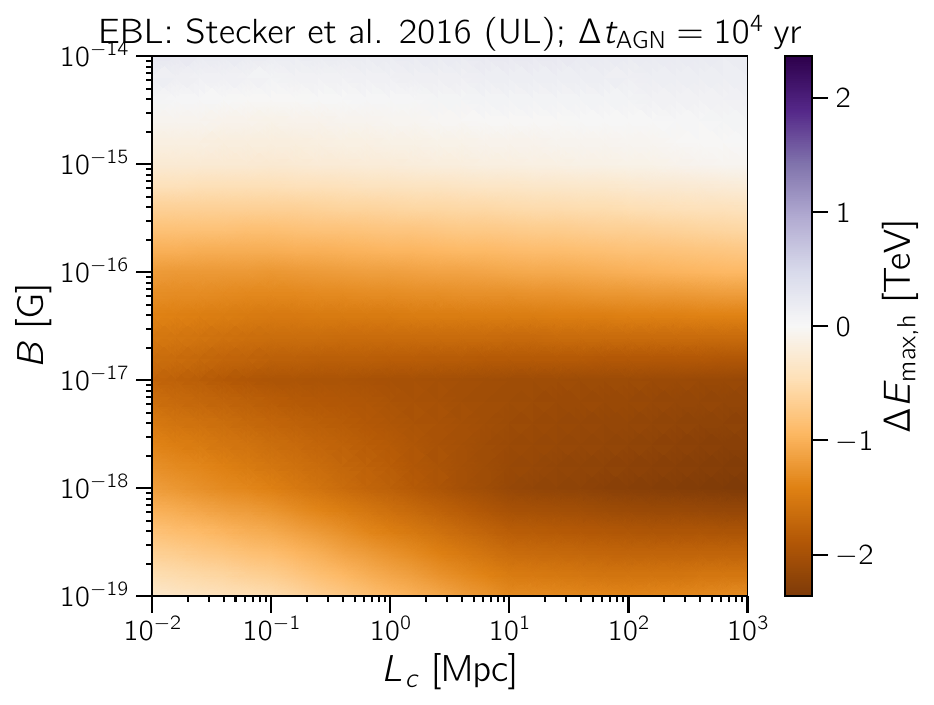}
  \caption{Average value of $\Emax$ (colour scale) for different combinations of $B$ and $\Lc$. Here $\Delta \Emaxh$ denotes the difference between the best-fit $\Emaxh$ for a given pair $(B, \Lc)$ and the corresponding quantity for $B=0$. The panels correspond to the indicated EBL models, assuming that \txs{} is active over $\dtAgn =10^4 \; \text{yr}$ in the low state.}
  \label{fig:emax}
\end{figure*}

As we have shown above, there are several combinations of $B$ and $L_{\rm c}$ that fit the given data well. However, this is only possible if for each of these pairs the correct intrinsic spectrum is chosen. Hence, one could reformulate it by saying that for each magnetic field configuration, there exists a preferred source model, i.e.~a set of spectral parameters $(\alpha_{\rm l},E_{\rm max,l},\alpha_{\rm h},E_{\rm max,h},\eta$). The change in the value of the maximal energy due to magnetic fields can be assessed by estimating the change in $\Emaxh$ with respect to the case without magnetic fields. This quantity will be henceforth called $\Delta\Emaxh$. Its average value, marginalized over all other quantities except $B$ and $\Lc$, is shown in figure~\ref{fig:emax}. From this figure it is clear that the actual value of $\Emaxh$ may change considerably, depending on the EBL model, if magnetic field effects are considered. The two bottom panels, corresponding to the lower and upper limit EBL model by \citet{stecker2016a}, respectively, point to an interesting trend: In the presence of IGMFs, compared to the $B=0$ case, the best-fitting $\Emaxh$ increases for \textit{all} considered magnetic field configurations in the case of the former EBL model, and decreases for \textit{all} combinations of $B$ and $L_{\rm c}$ for the latter one.

In addition, one can see that for three out of the four EBL models the change of $\Delta \Emaxh$ has a stronger dependence on the magnetic field strength ($B$) than on the coherence length ($\Lc$). This may be explained by the fact that, in general, time delays of electromagnetic cascades are predicted \citep{neronov2009a} to be independent of the coherence length for $\Lc \gg D_{e}$ (where $D_{e}$ is the electron energy loss length due to inverse Compton scattering) which, in turn, results in a smaller sensitivity of the estimates of the spectral parameters to the magnetic field.

The effects of the magnetic field on the spectral index were found to be tiny ($\lesssim 10^{-3}$) and are, thus, not shown. Moreover, the impact of $\dtAgn$ on the results is virtually negligible.

\section{General remarks}\label{sec:discussion}

A key aspect of this work concerns the injection of VHE gamma rays in the intergalactic medium by \txs. 
The fact that the combined observations by IceCube, MAGIC, and VERITAS \citep{icecube2018b,magic2018a,Abeysekara:2018oub} show a flux suppression above $E \gtrsim 400 \; \text{GeV}$ calls for an explanation. 
The first hypothesis is that this is due to EBL attenuation. The optical depth for a $\simeq 100 \; \text{GeV}$ photon for a source at $z=0.34$ is $\sim 0.1$, meaning that this hypothesis alone does not account for the data. Another possibility is that the absorption is due to the presence of a weak broad-line region~\citep{Padovani:2019xcv}. If this is true, then the emitting region of \txs{} would be located in the outer regions of the object, thus explaining why the spectrum extends up to $\sim 400 \; \text{GeV}$; otherwise, these VHE photons would be completely absorbed if they were to traverse a much larger region before escaping into intergalactic space. A third hypothesis attributes the cut-off to intrinsic absorption~\citep{keivani2018a,Petropoulou:2019zqp}. We suggest that the observed time dependence of gamma-ray fluxes from \txs{} is affected by intervening IGMFs. This, however, does not exclude any of the aforementioned possibilities.

We assumed that VHE gamma rays can indeed escape the blazar, which, as we argue below, is a valid approach, even though the details of the underlying mechanism are currently an active field of research. VHE gamma rays and neutrinos are associated to each other if they are created via processes that lead to the production of mesons (nucleus--nucleus and nucleus--photon interactions), whose decays are responsible for generating the observed particles. \citet{murase2016a} estimated the intrinsic optical depth ($\tau_{\gamma\gamma}$) for photons in the source to be $\tau_{\gamma\gamma} (E_{\gamma,0}') \sim 1000 \tau_{p\gamma}(E_\text{CR}')$, wherein $\tau_{p\gamma}$ is the corresponding optical depth for photomeson production by protons, the prime indicates that the quantities should be taken in the co-moving frame, and $E_{\gamma,0}' = 500 ( E_\text{CR}' / 100 \; \text{PeV})\; \text{GeV}$ (see also \citealt{mannheim2001a}). Here, the factor 1000 stems from the ratio between the $\gamma\gamma$ and $p\gamma$ cross-sections (cf.~also fig.~5 in \citealt{murase2016a} for an illustration, including some limiting cases).  Thus, it is reasonable to expect the gamma-ray spectrum to extend up to TeV energies -- although energies much higher than 1~TeV do represent a theoretical challenge. 

\citet{reimer2019a} claim that the photon fields within \txs{} required for the production of HE neutrinos would imply $\tau_{\gamma\gamma} \gg 1$ at $E_\gamma \sim 1 \; \text{GeV}$ energies and, as a consequence, neutrinos and GeV photons would not have the same origin. While this is indeed a plausible possibility, there is some freedom to change $\tau_{p\gamma}$ and $E_\text{CR}$ at the source. Moreover, we could easily evade this particular constraint by invoking cosmic-ray nuclei instead of nucleons. It suffices to demonstrate the plausibility of scenarios with $\Emax \gtrsim 1 \; \text{TeV}$ during the flaring state, and point out the limitations inherent to simple one-dimensional single- or few-zone models that neglect IGMFs, such that any conclusions relying on these simplified models ought to be extrapolated with caution.

The picture changes if we consider cosmic-ray nuclei instead~(see e.g.~\citealt{rodrigues2018a}). In this case, $\beta$-decays during nuclear cascades triggered by photodisintegration may provide a significant contribution to the neutrino flux and the main gamma-ray production channel (photomeson production) may become subdominant. One could also consider a neutral beam, as done by~\citet{zhang2020a}. In this case, however, the gamma-ray flux would be much lower compared to the neutrino one, given that pairs generated via Bethe--Heitler process would respond for a large fraction of the total gamma-ray flux.

\citet{padovani2018a} found that gamma rays from the object PKS~0502+049 contaminate the signal at energies $\lesssim \text{a few} \; \text{GeV}$. If a part of the flux measured by {\it Fermi}-LAT could, indeed, be attributed to this source, our results would change quantitatively. Nevertheless, this effect would be small, since only a few data points would be affected. 

The determination of the maximal energy attainable by cosmic rays in blazars such as \txs{} is an important issue, given its intrinsic connection with ultra-high-energy cosmic rays (UHECRs; $E \gtrsim 1 \; \text{EeV}$), whose origins elude us (see \citealt{alvesbatista2019d} for a review). This connection, in the context of blazars, has been discussed by~\citet{resconi2017a,rodrigues2018a,murase2014a,padovani2015a}. As pointed out by \citet{keivani2018a}, the X-ray emission in the $0.1 -- 100 \; \text{keV}$ band, combined with the HE and VHE data, seems to disfavour the possibility that this object would be an UHECR source, with $E_\text{CR} \gtrsim 1 \; \text{EeV}$. This connects with another issue regarding the classification of \txs{} as a BL~Lac or FSRQ (see \citealt{Padovani:2019xcv}). Neutrino production is more efficient in the latter, due to the rapid photodisintegration of nuclei~\citep{palladino2019a}, whereas the former may, indeed, accelerate cosmic rays to ultrahigh energies~\citep{rodrigues2018a,yoshida2020a}.

In light of our results, claims of correlations between HE neutrinos and HE/VHE gamma rays must be carefully made if the objects in question are not steady gamma-ray sources. For instance, \citet{kadler2016a} found a temporal correlation between the blazar PKS~1424+240 and neutrino events. This object is located at $z \simeq 0.60 -- 1.20$~\citep{rovero2016a}, so the VHE part of the gamma-ray flux is attenuated by the EBL and reprocessed to lower energies, potentially arriving many years after the emission, depending on the properties of the IGMFs \citep{neronov2009a}. Similar associations were reported by \citet{antares2012a}, whose results suggest coincidences between neutrino events and {\it Fermi}-LAT flaring blazars. Nevertheless, more recent analyses using a larger data set do not confirm the previous findings~\citep{antares2015a,ayalasolares2019a}. 
In these cases, like in the one we studied (\txs ), the actual contribution of cascade photons to the gamma-ray flux at the $\sim$~GeV--TeV band depends on the maximal energy and spectral index of the gamma rays escaping the source during the flaring activity, but if $\Emaxh \gtrsim 1 \; \text{TeV}$, this contribution may be dominant at $1--100$~GeV.

In our simulations we included energy losses due to synchrotron emission by electrons interacting with IGMFs, but we did not compute the associated spectrum. One could argue that this contribution could affect the X-ray part of the spectral energy distribution of \txs. While this is a valid concern, the total irradiated synchrotron power is very small, namely tens of orders of magnitude below measurements by Swift/NuSTAR~\citep{icecube2018b,keivani2018a}.

Our simulations were restricted to the case in which the blazar jet points exactly towards Earth, with no misalignment. Multiwavelength fits of the SED of \txs{} suggest a small misalignment angle of $\theta_\text{los} \simeq 0.8^\circ$~\citep{magic2018a}.
Moreover, to reduce the computational load we have neglected the jet opening angle of the object. In general, this angle is $\theta_\text{jet} \sim \Gamma^{-1}$, where $\Gamma$ is the bulk Lorentz factor of the jet. For \txs, this angle is estimated to be $\theta_\text{jet} \simeq 2.5^\circ$~\citep{magic2018a,keivani2018a,sahakyan2018a,gao2019a}.

Recent work by~\citet{halzen2019a} presents a phenomenological model similar to ours (see eq.~\ref{eq:spec}). The authors claim a successful description of the observations for $\Emaxl \simeq 100 \; \text{GeV}$ and $\Emaxh \simeq 1 \; \text{TeV}$, for $B = 10^{-19} \; \text{G}$ and $\Lc \simeq 1 \; \text{Mpc}$, analysing the 2014/2015 flare of \txs{} in their studies, together with the 10-yr {\it Fermi}-LAT flux that we also considered. These results are order of magnitude compatible with ours. However, we found stronger magnetic fields $B \simeq 10^{-14} \; \text{G}$ for $\Lc \sim 0.1--10 \; \text{Mpc}$ to provide better description of the data. This is not surprising, given that our analysis considers the 2017 flare and, in addition to {\it Fermi}-LAT data, takes into account the observations by MAGIC~\citep{magic2018a}. Furthermore, we have scanned a much broader range of magnetic field parameters, as shown in figure~\ref{fig:emax}, using detailed three-dimensional simulations.

Comprehensive multimessenger studies of blazar flares such as that of \txs{}, but with much larger samples, will enable us to properly infer the value of the cut-off energy ($\Emax$). Current observatories such as HAWC~\citep{hawc2017a}, as well as the upcoming Cherenkov Telescope Array~\citep{cta2019a} will have enough sensitivity in the 0.1--100~TeV region to provide temporal information at these energies. In particular, electromagnetic follow-ups of HE neutrino events from flaring blazars with real-time networks such as AMON~\citep{ayalasolares2020a} will play a key role in understanding VHE emission by blazars. 

Finally, as we have shown in \citet{saveliev2020b}, it is, in principle possible to constrain both the magnitude and the coherence length of IGMFs using multimessenger observations of neutrinos and gamma rays from objects such as TXS~0506+056. When new neutrino sources are observed, we will be able to apply this method to derive stricter limits on IGMFs.

\section{Summary and Outlook}\label{sec:conclusions}

We have here made the case for HE gamma-ray emission by \txs{} during the 2017 flare. We argued that the 1--100~GeV region of the measured flux may contain a significant contribution of cascade photons produced in electromagnetic cascades in the intergalactic medium. Some of these secondary photons may suffer from very large time delays and angular spreading due to the intergalactic magnetic field.

Our central thesis that the determination of the maximal intrinsic gamma-ray energy ($\Emaxh$) during the flaring state is influenced by IGMFs holds regardless of the actual mechanism responsible for the apparent flux suppression above $\gtrsim 400 \; \text{GeV}$. Nevertheless, there are theoretical arguments that make HE gamma-ray emission by blazars inefficient if neutrinos are efficiently produced. If this is the case, then $\Emaxh$ may be relatively low, and so can be the contribution of secondary cascade photons. Either way, it is worth considering that IGMFs may influence the inferred intrinsic spectral properties of \txs{} and other objects. Conversely, it will also be possible to constrain IGMFs using multimessenger observations of gamma rays and neutrinos.

In the future we plan to carry out a more thorough analysis, similar to this one, for other sources. Moreover, we intend to perform a more general combined analysis performing the usual SED fits using, for instance, one-zone models, \emph{in addition to} propagation effects due to IGMFs.

\section*{Data Availability}

The data underlying this paper will be shared on reasonable request to the corresponding author.

\section*{Acknowledgements}

We thank Maria Petropoulou for useful discussions. The work of AS was supported by the Russian Science Foundation under grant no.~19-11-00032, carried out at the Immanuel Kant Baltic Federal University. RAB gratefully acknowledges the funding from the Radboud Excellence Initiative, and from the S{\~a}o Paulo Research Foundation (FAPESP) through grant \#2017/12828-4 in the early stages of this work. Part of the simulations were performed in the computing facilities of the GAPAE group at Institute of Astronomy, Geophysics and Atmospheric Sciences of the University of S{\~a}o Paulo, FAPESP grant: \#2013/10559-5.





\label{lastpage}
\end{document}